\documentclass[aps,pre,preprint,groupedaddress,showpacs]{revtex4}

\usepackage{graphicx}
\begin{document}


\title{A Novel Approach Applied to the Largest Clique Problem}


\author{Vladimir Gudkov }
\email[]{gudkov@sc.edu}
\affiliation{Department of Physics and Astronomy
\\ University of South Carolina \\
Columbia, SC 29208 }
\author{Shmuel Nussinov}
\email[]{nussinov@ccsg.tau.ac.il}
 \affiliation{
Department of Physics\\
Johns Hopkins University \\
Baltimore MD 21218\\
and\\
Tel-Aviv University, School of Physics and Astronomy \\ 
Tel-Aviv, Israel \\
}
\author{Zohar Nussinov}
\email[]{Zohar@lorentz.leidenuniv.nl}
\affiliation{Institute Lorentz for Theoretical Physics, Leiden University\\
POB 9506, 2300 RA Leiden, The Netherlands\\
}


\date{\today}

\begin{abstract}
A novel approach to complex problems has been previously applied to
graph classification and the graph equivalence problem.  Here we
apply it to the NP complete problem
of finding the largest perfect clique within a graph $G$.
\end{abstract}

\pacs{89.75.Hc, 89.90.+n, 46.70.-p, 95.75.Pq}


\maketitle



\section{Introduction}

In a novel dynamical approach the $n$ vertices of a graph $G$ are mapped onto $n$      physical points located initially at equal distances from each other forming a symmetric $n$  simplex in $n-1$ dimensions.  Attractive/repulsive forces are introduced between pairs of points corresponding to connected/disconnected vertices in the original graph $G$.  We then let the system evolve utilizing first order Aristotelian dynamics\cite{gjn,gn,nn}. We tune the relative strength of repulsive and attractive forces to be $v/n$ with  $v$ the average valency i.e. average number of vertices connected to a given vertex so as to have no net average repulsion/attractions. 

We found, that as the system evolves  various physical clusters of points tend to form.  These physical clusterings reveal clusters (or imperfect cliques) in the graph  -  namely groups of vertices with a larger than average mutual connectivity.  Also the matrix of distances $R_{ij}(t)$  between the various points $\vec{r}_i(t)$    and $\vec{r}_j(t)$ is characteristic of the graph topology: points corresponding to vertices which are ``close in the graph'' namely have (relatively) many, short, paths connecting them will move closer together and conversely, points which are ``far in the graph'' tend to move apart.  

The distance matrix and clusters are important graph diagnostics.  In particular the first allows us to solve easily the graph equivalence problem namely to decide if two connectivity matrices $C_{ij}$ and $C\prime _{ij}$ correspond to the same topological graph and if they do to find the relabelling of vertices which makes $C$ and $C\prime$ identical.  

These results are of considerable practical importance. Still neither of the above problems belongs in the special class ``NP complete'' problems.  The latter consists of problems such as the travelling salesmen problem and the satisfiability  problem for which a putative solution can be readily checked in polynomial time yet no polynomial solution method is presently known\cite{comp}.  

Many of these problems can be phrased in terms of graphs as the task of finding  some specific graph  $g$ inside bigger graph $G$.
Further, all these problems which superficially seem very different are at a basic level ,equally difficult: If a method of solution in polynomial time is found for one such problem then all the problems should be solvable in such time by essentially the same method. Conversely if we can prove that just one NPC problem necessarily require, non-polynomial  time for its solution, the same holds for all of them.  
Two of us have recently conjectured\cite{nn} that a new variant of our approach namely of dynamically docking rigid simplexes $s$ and $S$ representing $g$ and $G$ can solve the ``g inside G'' problems.
Here we wish to present the first concrete application of the original, point translation or single simplex distortion algorithm (SDA), to an NPC problems namely that of finding the largest perfect clique in $G$.

To most clearly illustrate the essence of the problem we consider the ``students in dorm'' example used in the general description of the Clay institute prize offered for resolving $P =NP$ problem\cite{web}.   We have $N=400$ students out of which we need to select $n=100$ which can live together in a dorm, subject to a very long list of mutual exclusions.  This list states that student \# 1 cannot be together with any one student from a specific set of  say 200 other students, student student \# 2 cannot be together with any one from another partially overlapping set with a comparable number of students etc. 
How can we pick up a set of 100 students such that any one is completely compatible with the other 99, and what is this set?  Clearly this is a particular example of the general satisfiability problem where the conditions imposed are just "two body" exclusions.

It is also a particular case of looking for a graph $g$ inside $G$ where $g$ is a perfect clique of vertices each of which is connected to all other members in the clique.  We encode into $G$ with $N=400$ vertices the various mutual exclusion constraints by  not connecting with edges vertices $V_i$ and $V_j$ if student \# i and student \# j are not compatible, and connecting by edges compatible pairs.  Clearly if we find within $G$ a clique with $n$ vertices it means, by our very construction, that the students to which these vertices correspond are indeed all mutually compatible.
We could  construct in judicious manner various smaller consistent subsets, and try piece them together.  Often, however a new inconsistency is revealed and we need to pursue other alternatives.  While we certainly can do this in far less steps than $\left(%
\begin{array}{c}
  N \\
  n \\
\end{array}%
\right) =\left(%
\begin{array}{c}
  400 \\
  100 \\
\end{array}%
\right)$ the difficulty of the problem seems to grow at least  exponentially with $n$.                

In desperation we might decide to resort to the following primitive alternative and simply let the 400 students ``fight it out''.  In this all out war each student will try to push away members which are inconsistent with him and pull in those which are.  This collective natural selection of the ``compatible'' - which may well be a prerelevant social phenomena - would hopefully leave us with the desired large group of mutually consistent individuals.  Unfortunately  the outcome of such a 400 way ``Somo'' fight of staying in the ring is strongly biased by the initial arbitrary placement of students in the two dimensional arena\cite{fn1}.  Thus we could envision a situation  where an ideal group of completely compatible dorm candidates is placed in  the center of a group of highly unpopular ones and is ``ejected'' together with them.
In order to generate the correct large clique we need to completely unbias the starting position and avoid the severe constraints due to our existence in a physical world with limited number of dimensions.
This can be done only if we go to $d=N-1$ dimensions and place the ``students'' which, in the inverse problem that we are really after, are  metaphors for the physical points representing the $N$ vertices of the graph,  at the vertices of a symmetric $N$ simplex.

\section{Searching for Cliques}

\begin{figure}[h]
\includegraphics{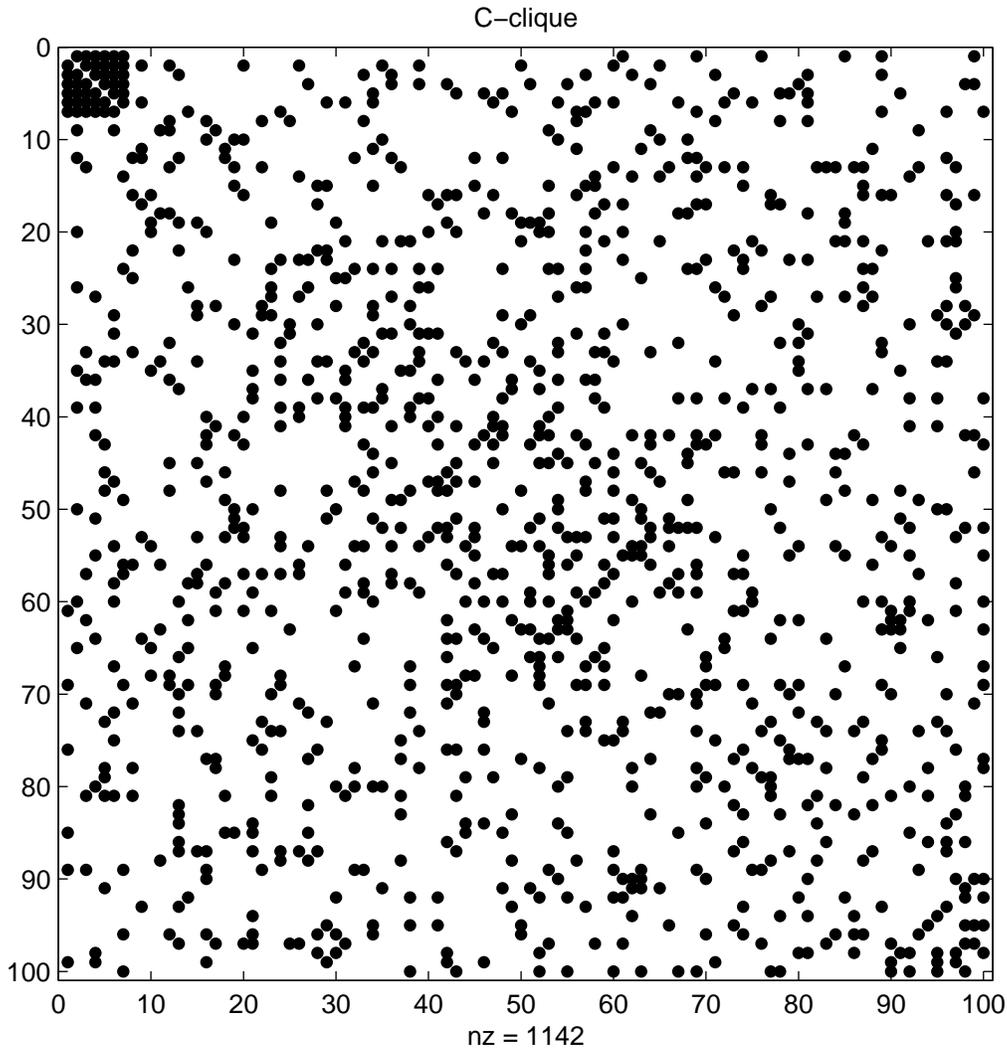}
\caption{Connectivity matrix $C$ with $7\times 7$ clique.}
\label{fig:clq-C}
\end{figure}

\begin{figure}[h]
\includegraphics{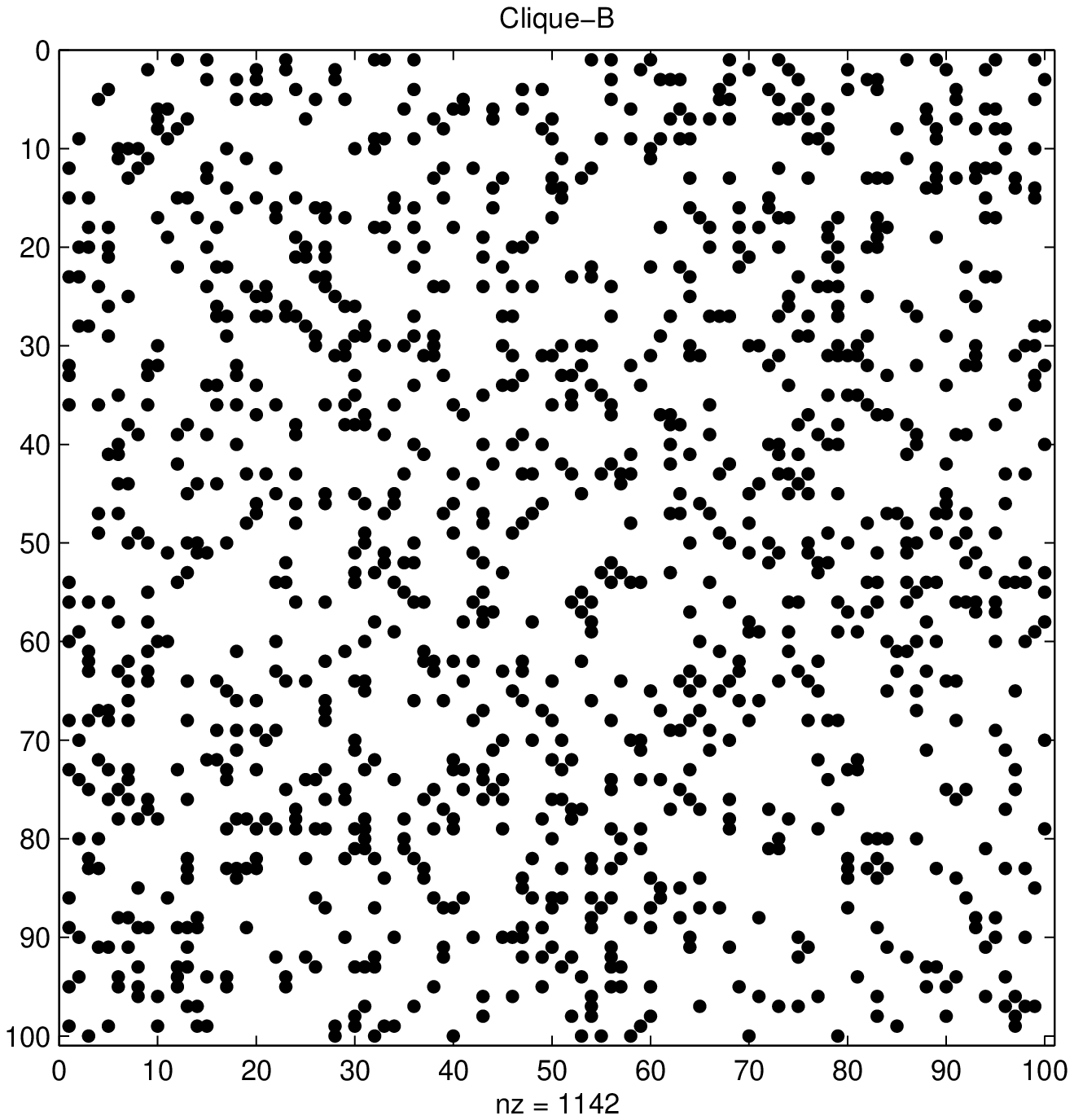}
\caption{Reshuffled connectivity matrix with $C\prime$ $7\times 7$ clique.}
\label{fig:clq-B}
\end{figure}

\begin{figure}[h]
\includegraphics{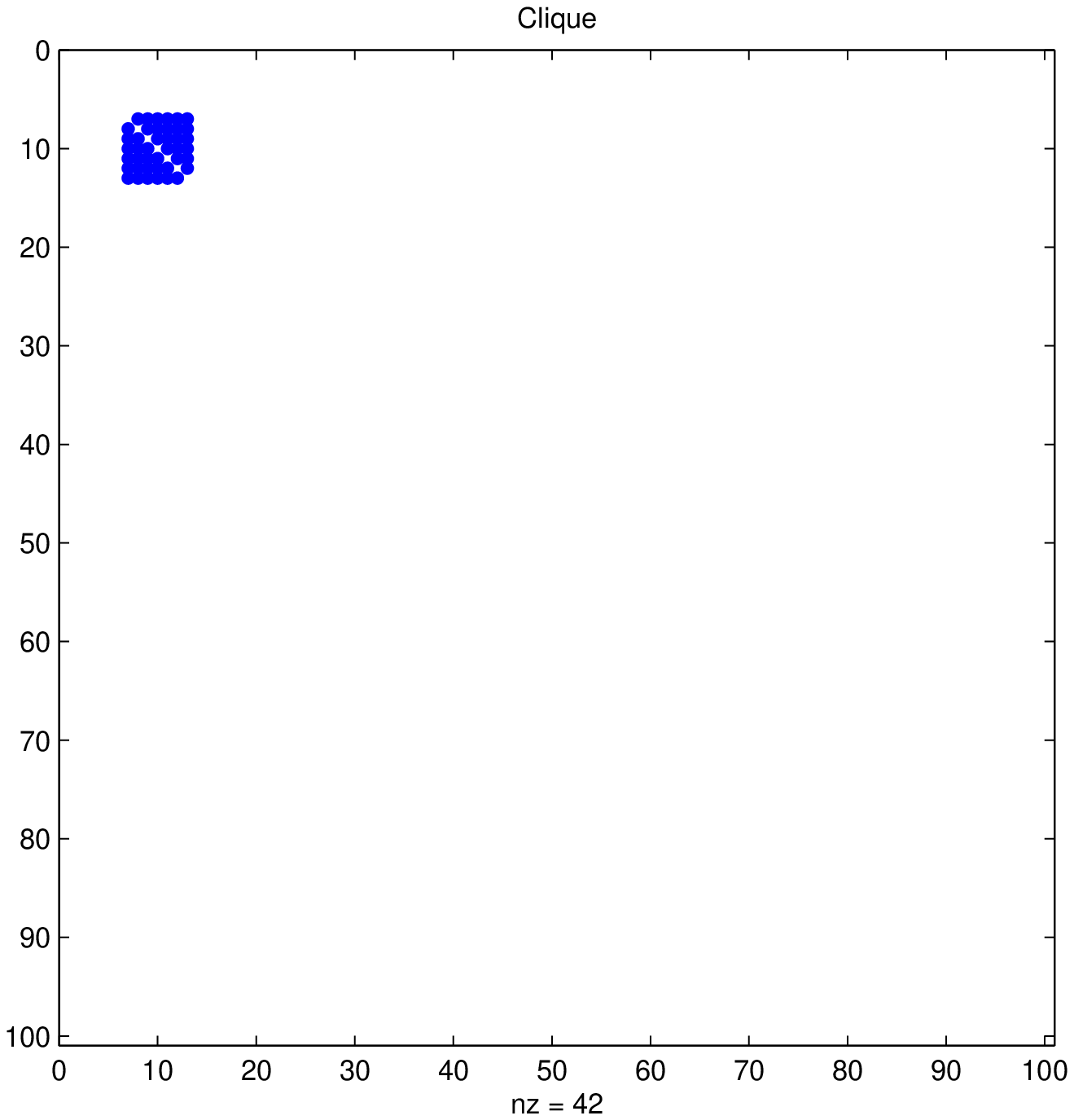}
\caption{Clique connectivity matrix for reshuffled connectivity matrix $C$.}
\label{fig:clq}
\end{figure}

  Our search for perfect cliques uses the same physically motivated
   dynamical algorithm previously developed to  identify  via the
   physically  bunched points clusters or ``imperfect cliques'' in a
   graph\cite{gjn}.
   We found that to adapt this algorithm for the present purpose we
   need only to enhance the ratio of the
   repulsive and attractive interactions.
   Originally it was chosen to be:
   \begin{equation}
   \label{x}
   U_{rep}(r)/U_{att}(r) = v/n,        
\end{equation}
   which could be relatively small. Thus for an average valency of $10$
   in a graph with $100$ vertices it is only $0.1$.
    However, in order to meet the criteria of {\em perfect} cliques we clearly
    have to significantly enhance the strength of the repulsive
    interactions so as to avoid points which are connected to a fairly
    large number of the points in the clique but not to ALL of them
   from joining in.
    Thus in the first round of applications we used
 \begin{equation}
   \label{xcl}
   U_{rep}(r)/U_{att}(r) = 1.        
\end{equation}
    We first considered a small clique of $n=7$ in a graph with $N=100$
   vertices with the connectivity matrix of Fig.(\ref{fig:clq-C}).
 In addition to the clique this matrix consists of six clusters with randomly created internal connections with average valency $20\%$. These clusters, in turn, have been randomly interconnected with a large valency $10\%$.  To simulate a real-life situation of networks with unknown structure (topology) we randomly permute the rows and columns of the matrix $C$ obtaining the reshuffled matrix $C^{\prime}$ shown in the Fig.(\ref{fig:clq-B}). 
Next we apply our algorithm for clusters reconstruction using equal attractive and repulsive constant forces  in $n-1=99$ dimensional space. The vertices of the 100-simplex were allowed to move under the influence of the forces on the 98-dimensional hyper-sphere in 99-dimensions. After a number of steps we analyzed the mutual distances between the vertices of the simplex and group neighbors which are close to each other into cliques. The new cluster-connectivity matrix is shown in Fig.(\ref{fig:clq}).
We see that due to the large repulsive forces most vertices did not move close each to others. The only vertices grouped together are the ones that belong to the clique.  

\begin{figure}[h]
\includegraphics{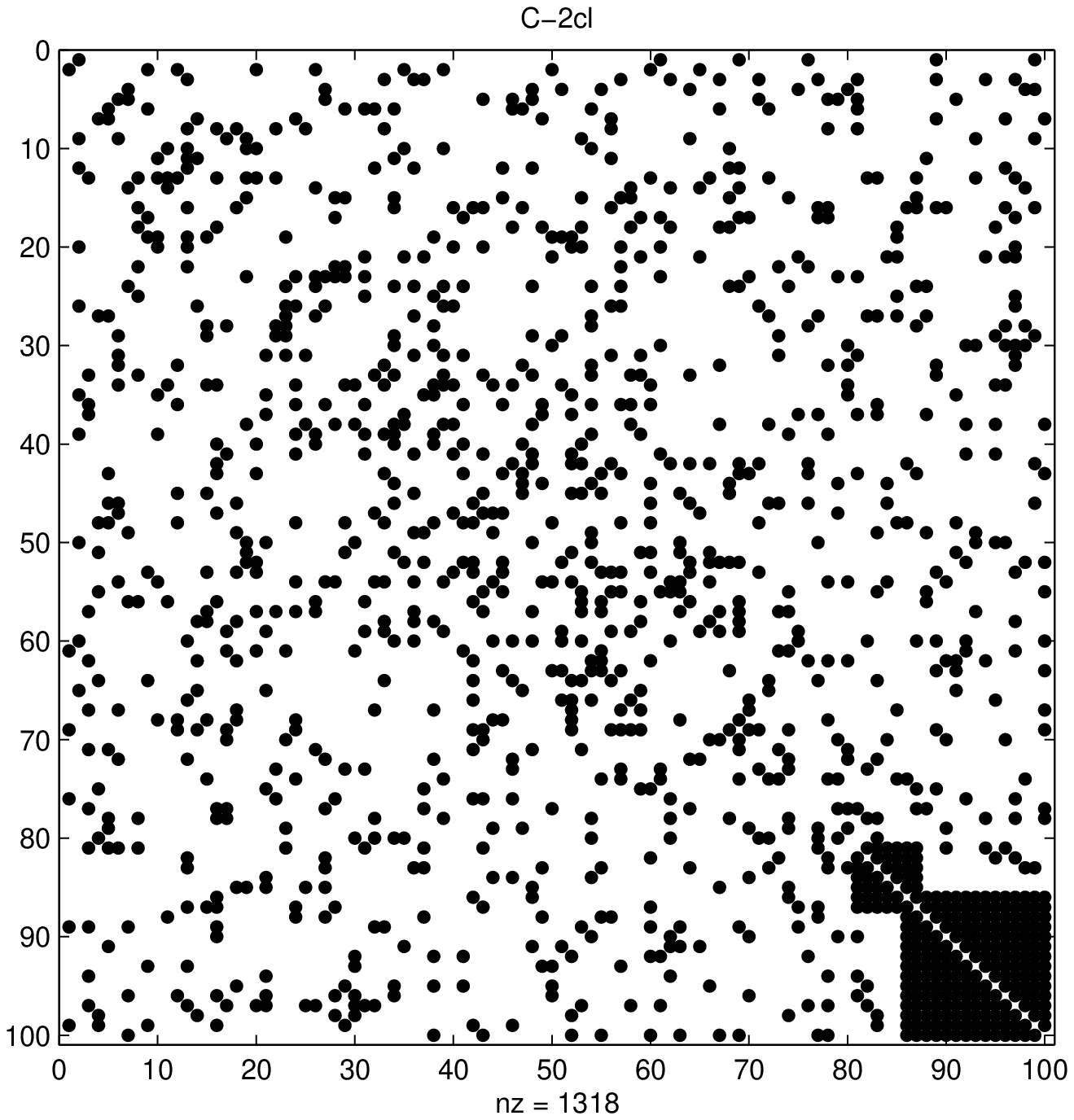}
\caption{Connectivity matrix $C$ with $7\times 7$ and $15\times 15$ cliques.}
\label{fig:clq2-C}
\end{figure}

\begin{figure}[h]
\includegraphics{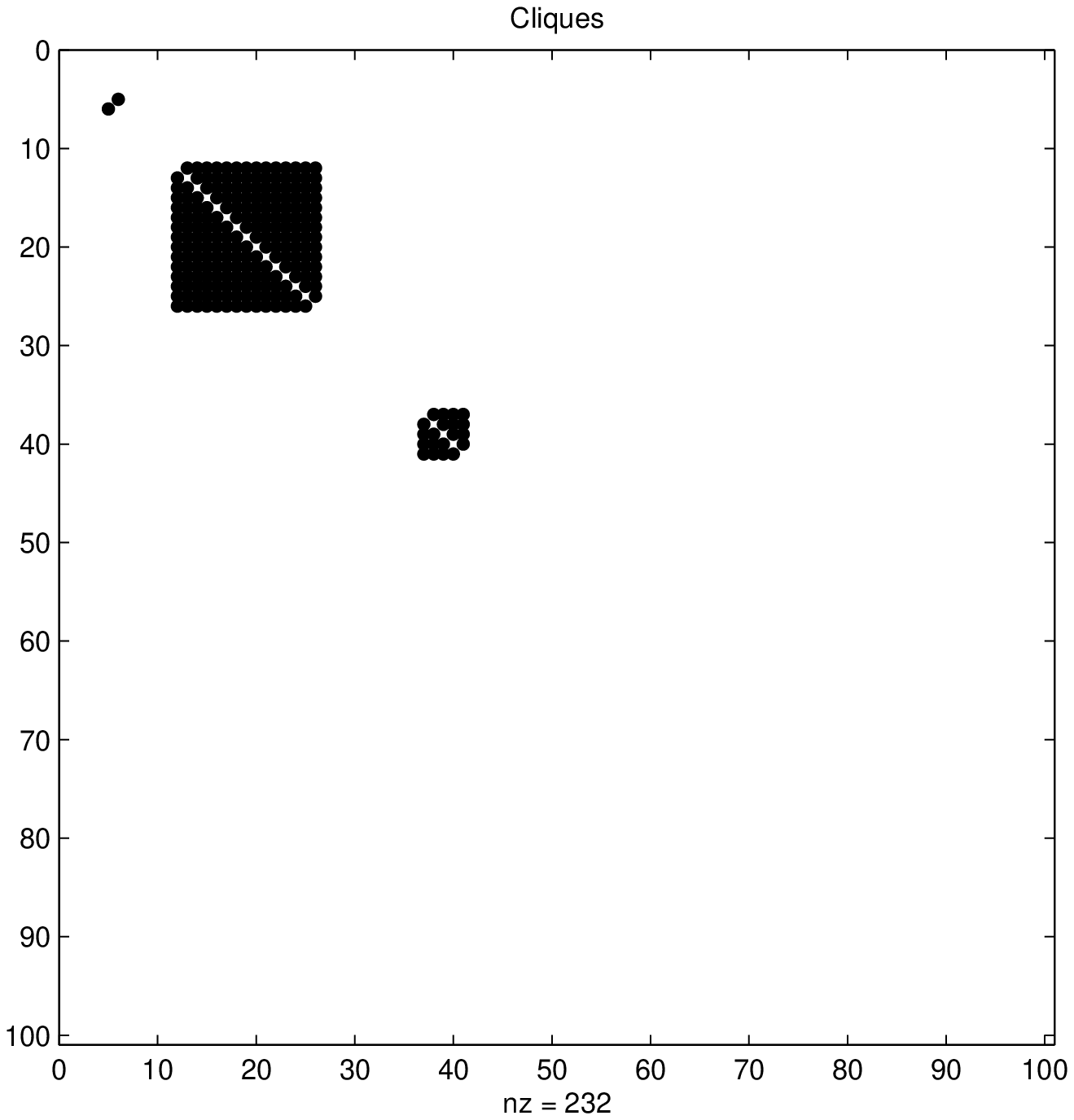}
\caption{Reconstructed clique connectivity matrix for $C\prime$ $7\times 7$ and $15\times 15$ cliques.}
\label{fig:clq2-cl}
\end{figure}

\begin{figure}[h]
\includegraphics{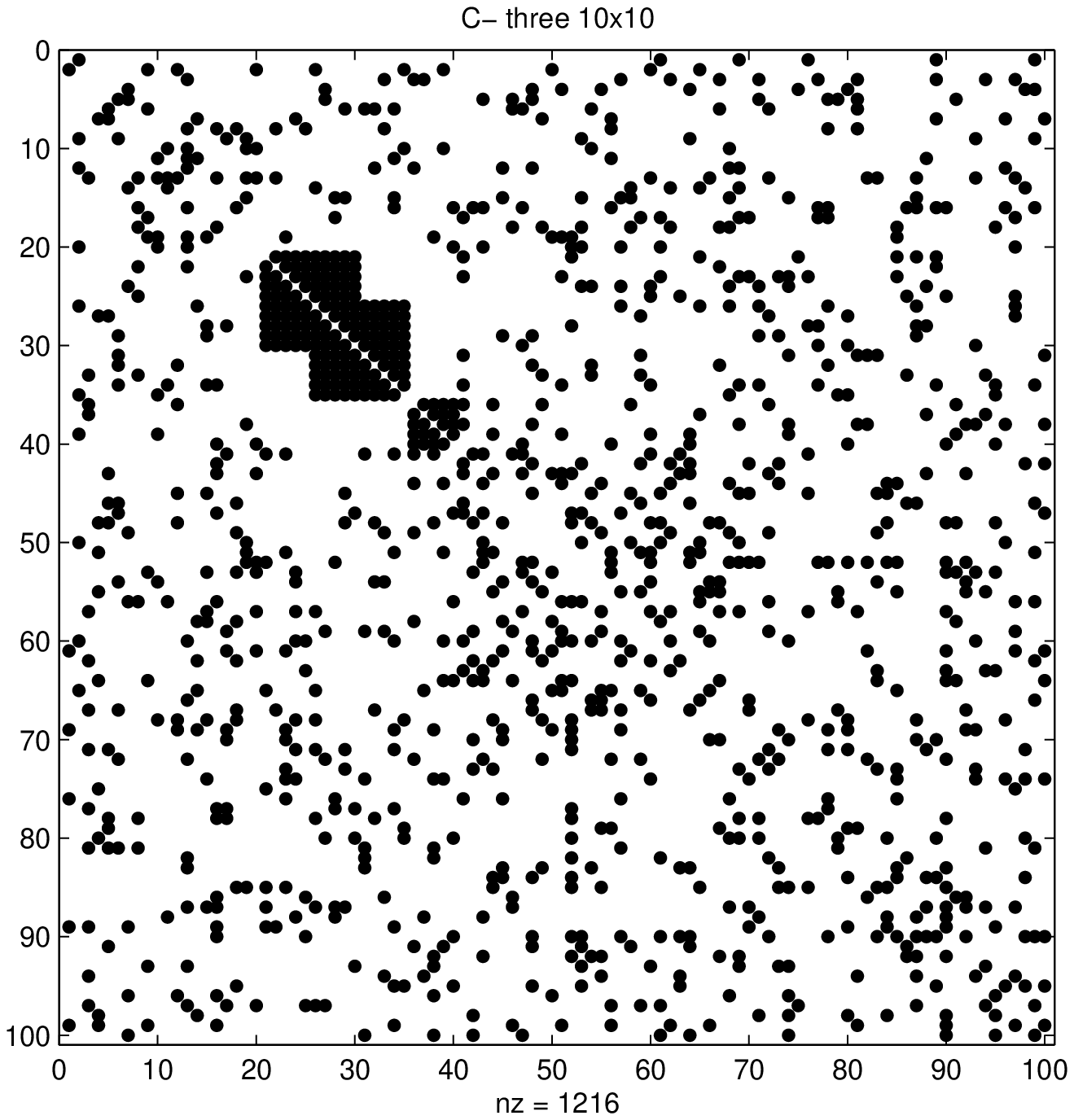}
\caption{Connectivity matrix $C$ for two  $10\times 10$ overlapped cliques.}
\label{fig:clq3-C}
\end{figure}

\begin{figure}[h]
\includegraphics{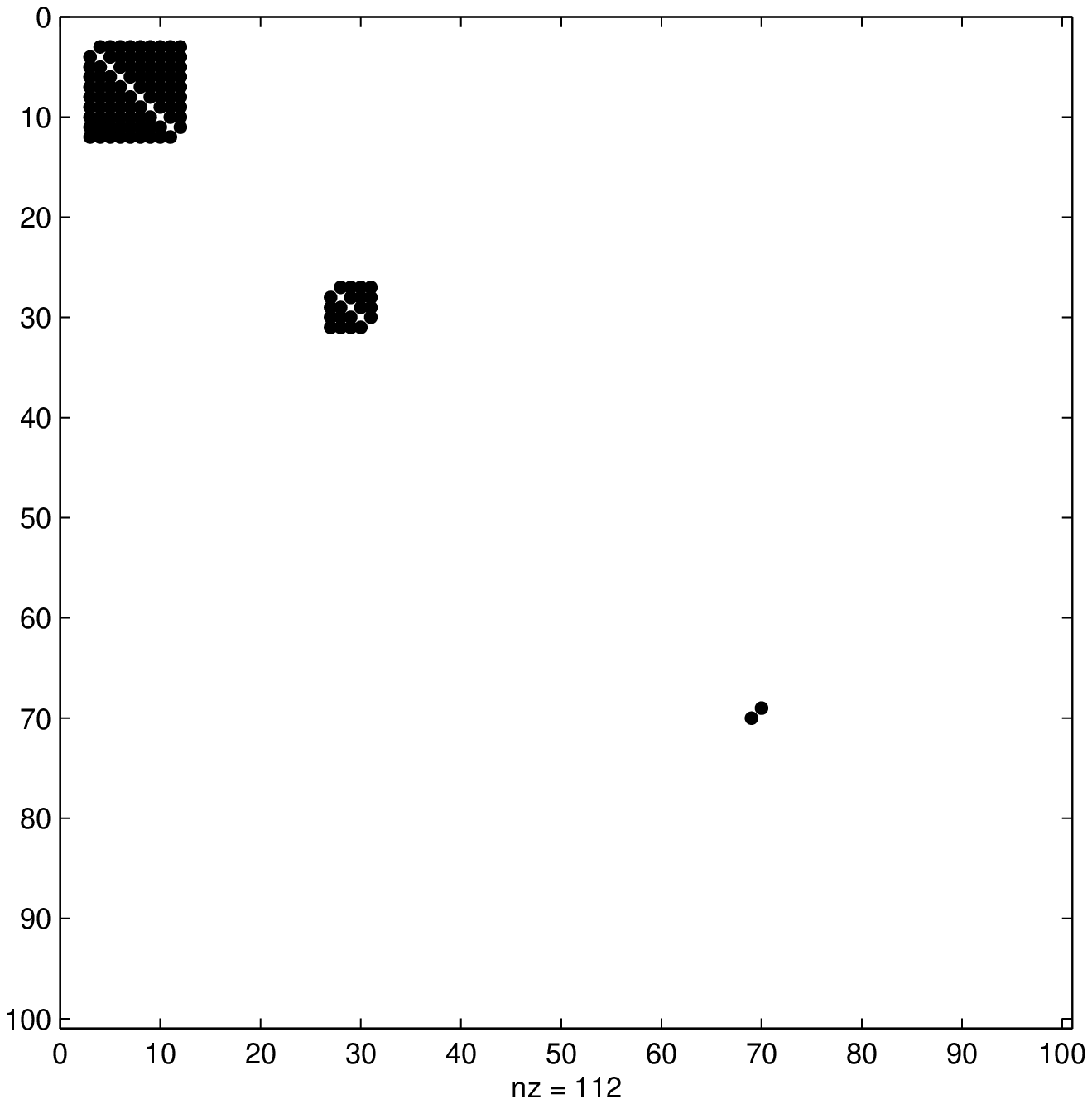}
\caption{Reconstructed clique connectivity matrix for two  $10\times 10$ overlapped cliques.}
\label{fig:clq3-cl}
\end{figure}

To see how the algorithm works for the case of overlapping cliques we considered two examples. The first example includes two cliques $7\times 7$ and $15 \times 15$ with a  $2\times 2$ overlap on a ``background'' of a $100\times 100$ matrix with the same average $10\%$ connectivity as above. The corresponding connectivity matrixes before  reshuffling is shown in Fig.(\ref{fig:clq2-C}). 
The reconstructed connectivity matrix for the cliques is shown in Fig.(\ref{fig:clq2-cl}).
The second example involves two of cliques $10$  with a large overlap of $5$ in Fig.(\ref{fig:clq3-C}). Our reconstruction yields one clique of $10$ and one of $5$ Fig.(\ref{fig:clq3-cl}). As expected we fully reconstruct the largest cliques. This is done at the expense of correspondingly reducing the size of  the reconstructed part of the overlapping smaller or equal size cliques.

\begin{figure}[h]
\includegraphics{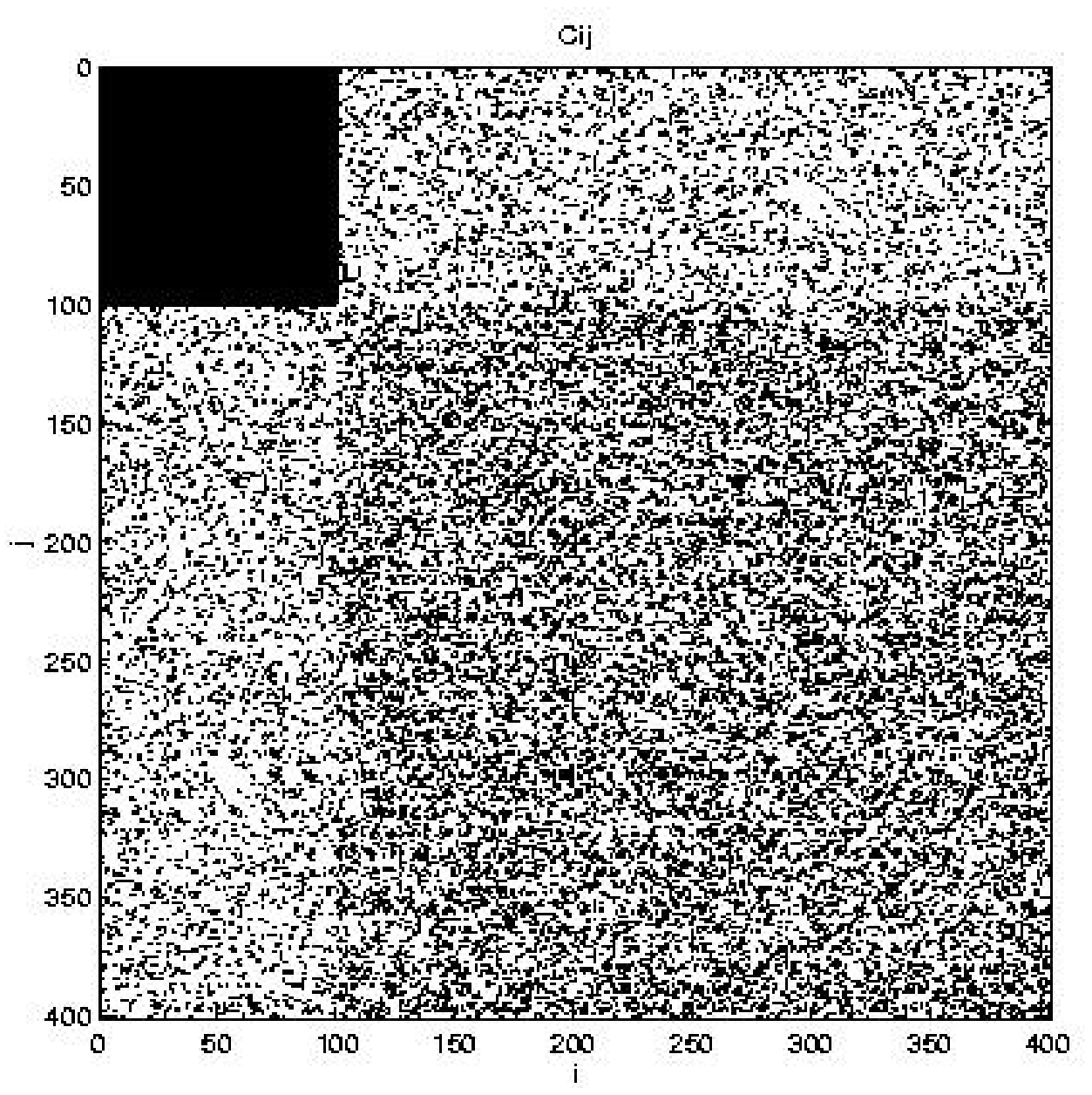}
\caption{Connectivity matrix $C$ for $100\times 100$ clique.}
\label{fig:clqnp-C}
\end{figure}

\begin{figure}[h]
\includegraphics{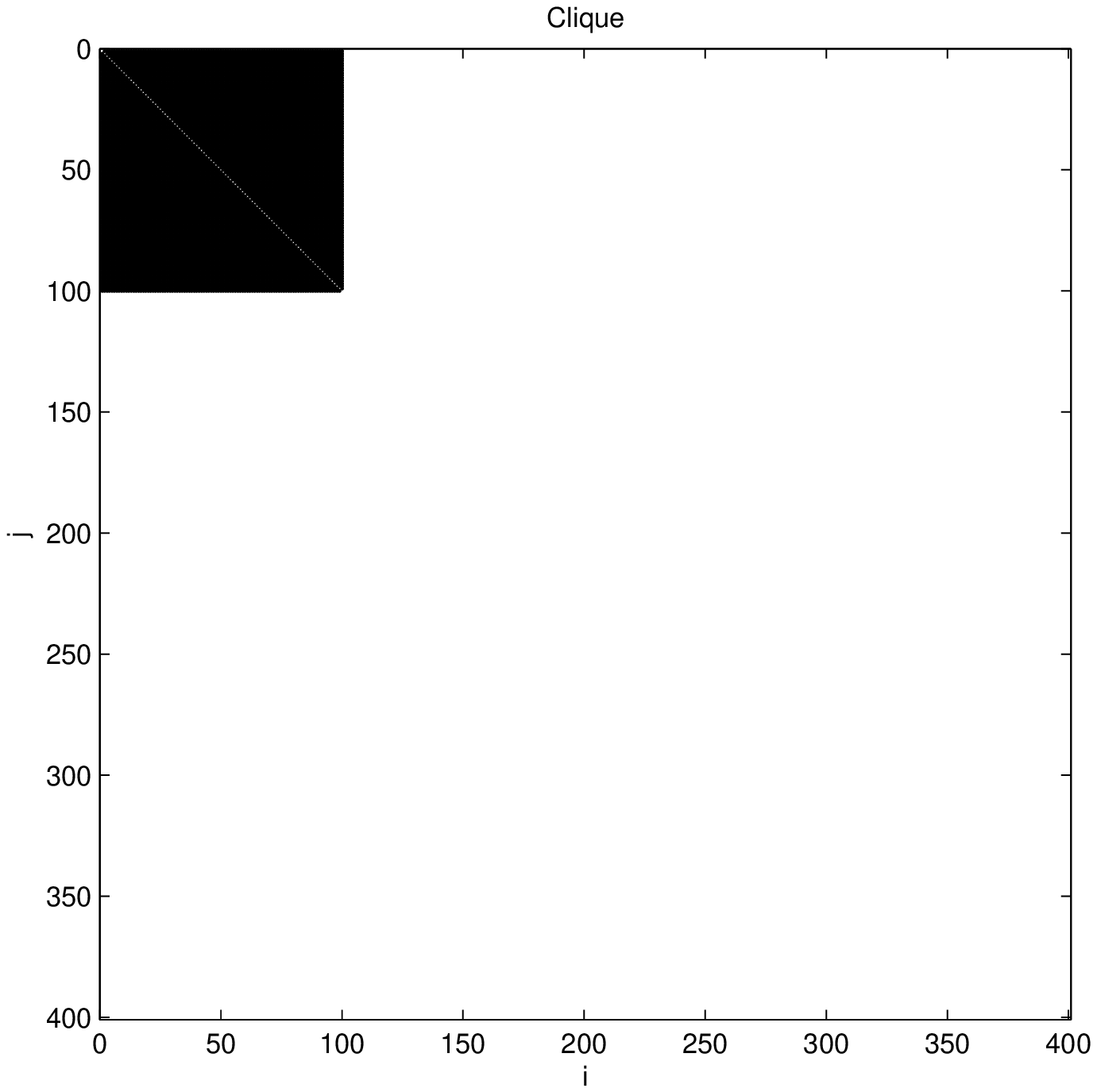}
\caption{Reconstructed clique connectivity matrix for $100\times 100$ clique.}
\label{fig:clqnp-cl}
\end{figure}

\begin{figure}[h]
\includegraphics{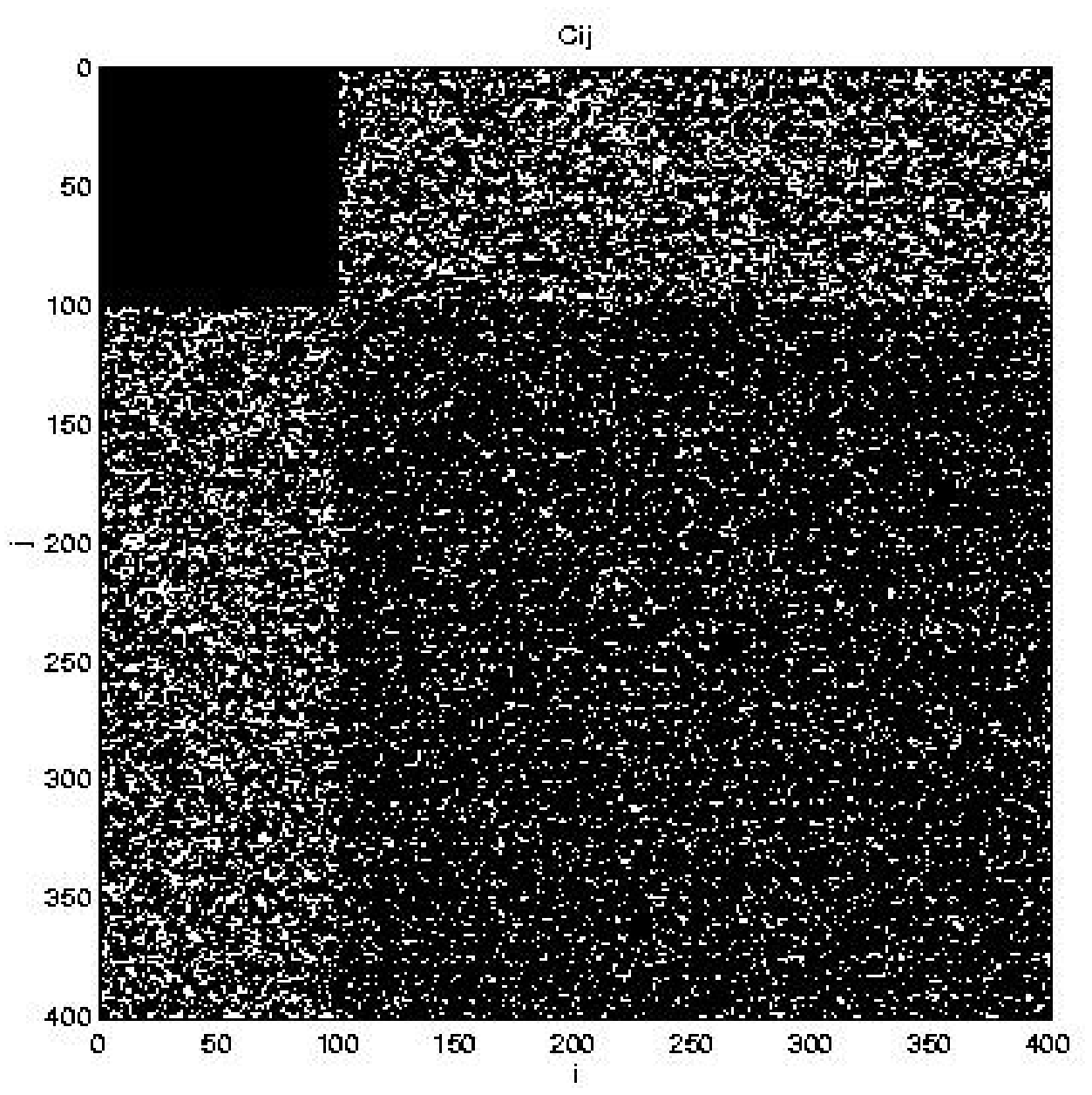}
\caption{Connectivity matrix $C$ for $100\times 100$ clique with ``heavy'' background.}
\label{fig:clqnp3-C}
\end{figure}

\begin{figure}[h]
\includegraphics{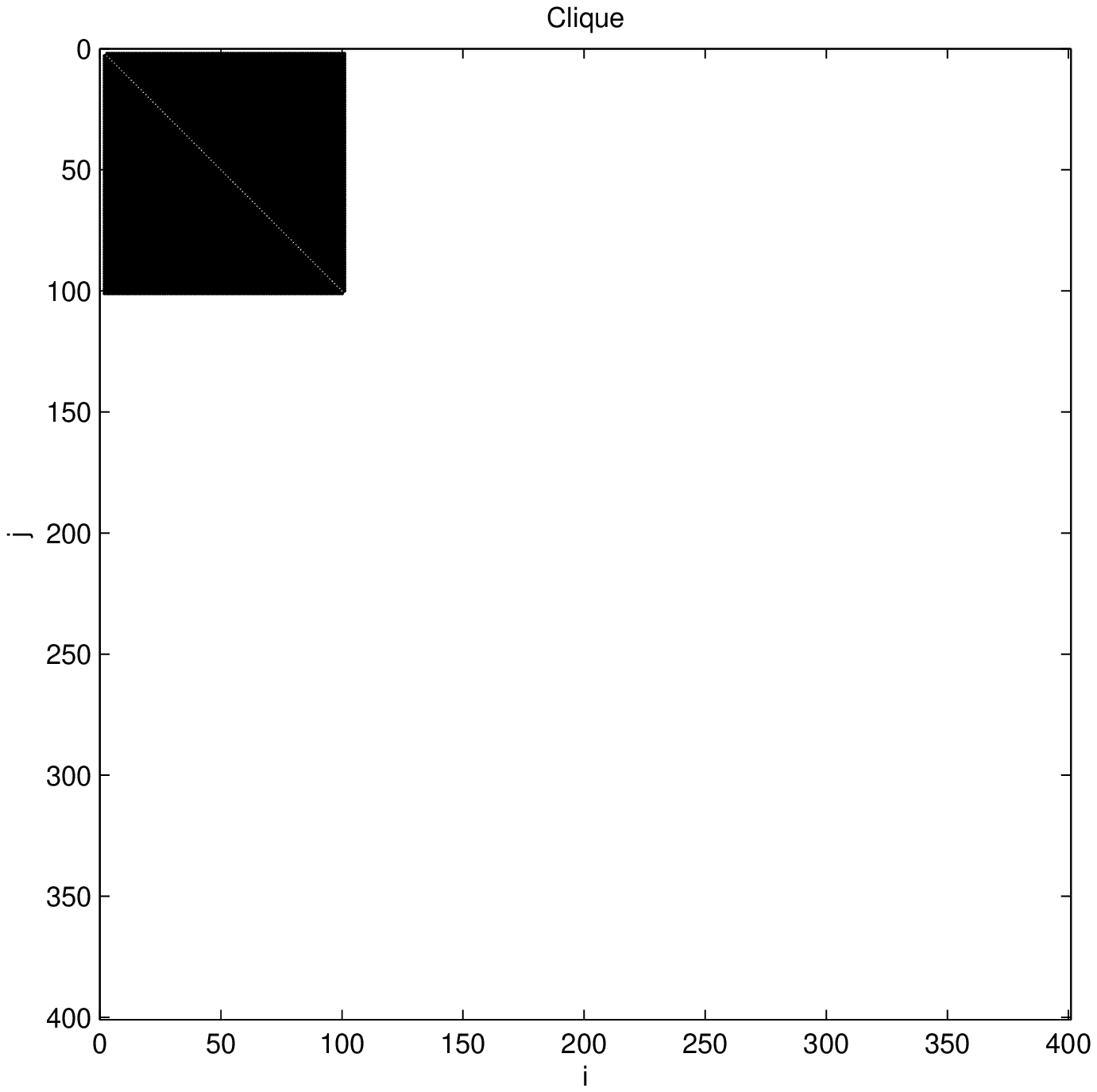}
\caption{Reconstructed clique connectivity matrix with doubled repulsive forces for $100\times 100$ clique  with ``heavy'' background.}
\label{fig:clqnp3-cl2}
\end{figure}

Other examples involve a $n=100$ clique in  a $N=400$ graph corresponding to the ``students in dorm'' question. In addition we had an imperfect clique or cluster  of $300$ with average valency of $20\%$ on a background of $10\%$ (Fig.(\ref{fig:clqnp-C})). The successfully reconstructed $100$ clique after reshuffling is shown in Fig.(\ref{fig:clqnp-cl}). 
 We next increase the ``background'' of average valency to $90\%$  for the $300\times 300$ cluster and  to $50\%$ of that average interconnections in Fig.(\ref{fig:clqnp3-C}).  
\begin{figure}[h]
\includegraphics{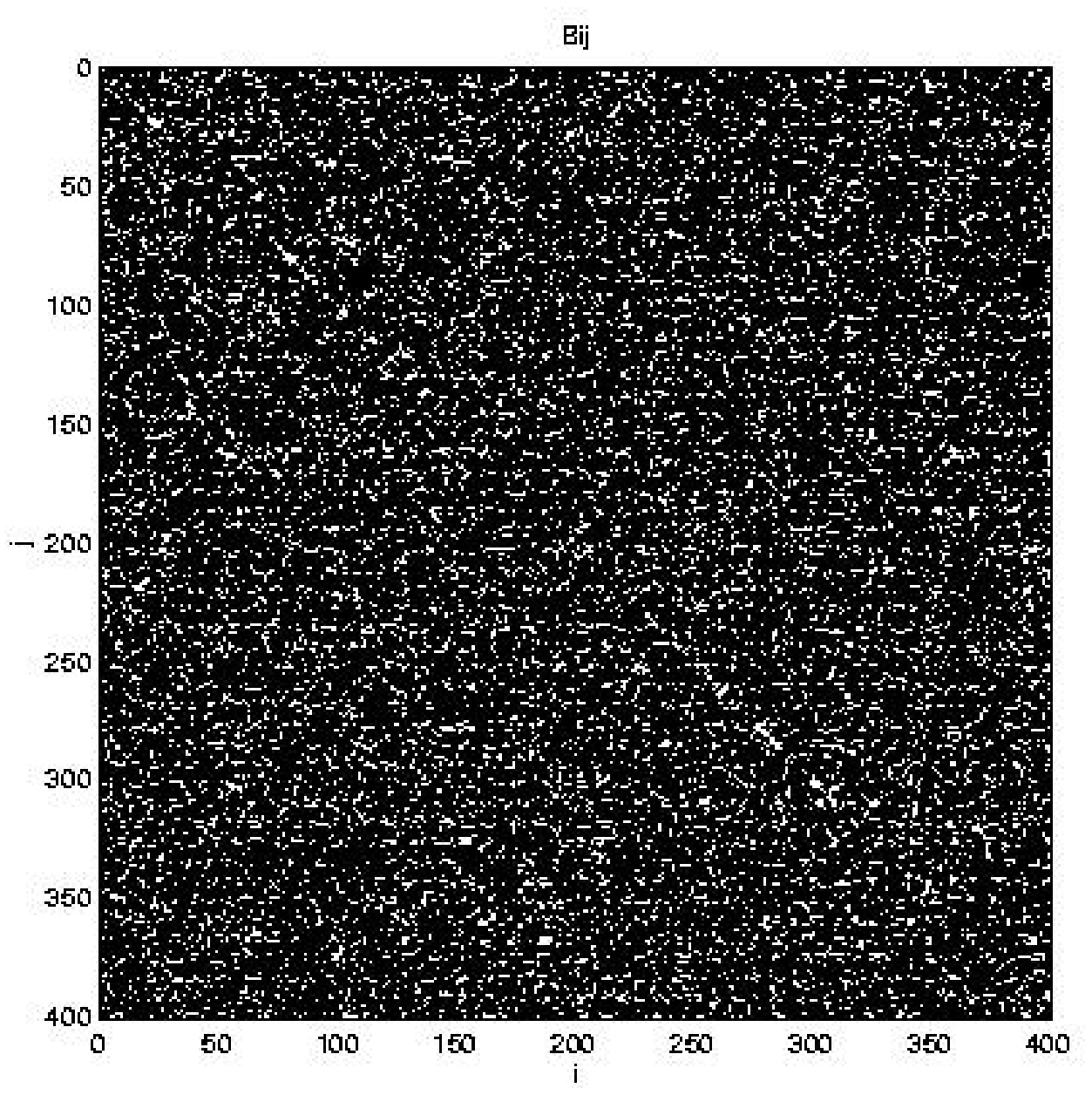}
\caption{Reshuffled connectivity matrix for $100\times 100$ clique with ``heavy'' background.}
\label{fig:clqnp3-B}
\end{figure}
\begin{figure}[h]
\includegraphics{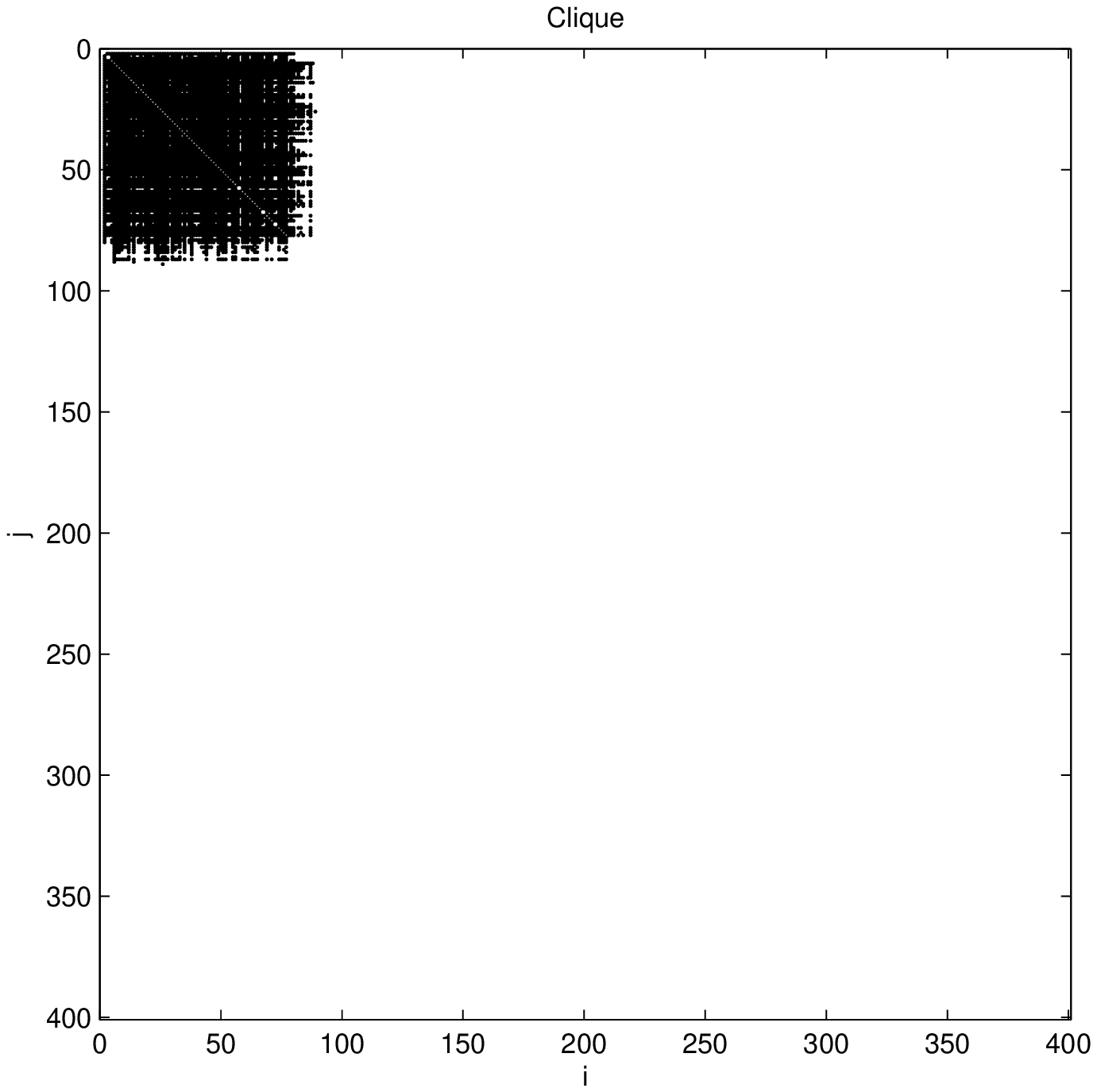}
\caption{Reconstructed clique connectivity matrix for $100\times 100$ clique  with ``heavy'' background.}
\label{fig:clqnp3-cl}
\end{figure}
Fig.(\ref{fig:clqnp3-cl}) shows the reconstructed results after reshuffling. The clique is imperfectly reconstructed. However, increasing $U_{rep}/U_{att}$ from one to two improves the reconstruction as shown in Fig.(\ref{fig:clqnp3-cl2}).

 The above results are to our mind fairly impressive. These results show that
    by mild tuning our original code solves in very short time and for
    many cases the NPC problem of the largest clique.
    It may still fall short of solving it in all cases.
     As a  worst case scenario we could envision a vertex ( or
    several such vertices) which are connected to all the vertices in
   the clique save one.
   To avoid these vertices from joining the clique even in this case,
  rendering it imperfect, we need that the single 
   repulsion due to the missing edge, overcome all the $n-1$ attractions
   to the rest of the points in the clique .
   Thus the strict perfect clique worst case scenario demands
\begin{equation}
\label{y}
U_{rep}> (n-1)\cdot U_{att}. 
\end{equation}
 This wildly differs from the above eq.(ref{x}): for a graph $G$ with $100$
  vertices $v = 10$ and a clique of size $n=10$ we need a factor hundred
  enhancement of the  ratio $U_{rep}/U_{att}$ from $0.1$ to $10$!

   Our $3-d$ based intuition would strongly suggest that this 
    stops formation of all cliques, perfect or not, since as any
  given point tries move towards its ``Designated'' clique it may be
  ``Overwhelmed'' by the many repulsive forces which will prevent it
  from joining the clique. The configuration with the perfect
  clique (and the largest perfect clique in particular) fully formed
  i.e having all its vertices collapse at a point is indeed the desired
  final lower energy state. However there may be false local minima
  which trap our system just like in spin glass\cite{spin} and protein folding problem\cite{bio}.
  
  This is indeed most certainly the case for "low" dimensionalities.
  However with $d= N-1$, as is the case here, the above intuition fails.
   Specifically  any one given ``test point'' feels just as many
  different forces in the directions of the other particles namely $N-1$
  as there are independent directions $d=N-1$ to move in.
  Ideally therefore the test particle should be able to
  simultaneously  respond to all different $N-1$  forces, move in the
  direction of all the attractors  and away from all the repellers
  and in the process further lower the energy of the system.
  We can adopt a local, non-orthogonal, system of  coordinates where the
  $N-1$ axes are aligned along the unit vectors pointing from $\vec{r}$ - the chosen
  point, to $\vec{r}_1,  \vec{r}_2, \ldots \vec{r}_{N-1}$ the other $N-1$ points.
  Using our choice of constant forces\cite{fn2} we have then
  a net force
\begin{equation}
\label{z}
 F(\vec{r})= \sum_{i=1}^{N-1}\frac{\vec{r}-\vec{r}_i}{|\vec{r}-\vec{r}_i|},
\end{equation}
 which is the sum of the unit vectors along these axes with + and - signs.
 Since these are $N-1$ linearly independent  vectors the sum never
  vanishes $|F(\vec{r})|>0$ always and no local minimum arises.

\begin{figure}[h]
\includegraphics{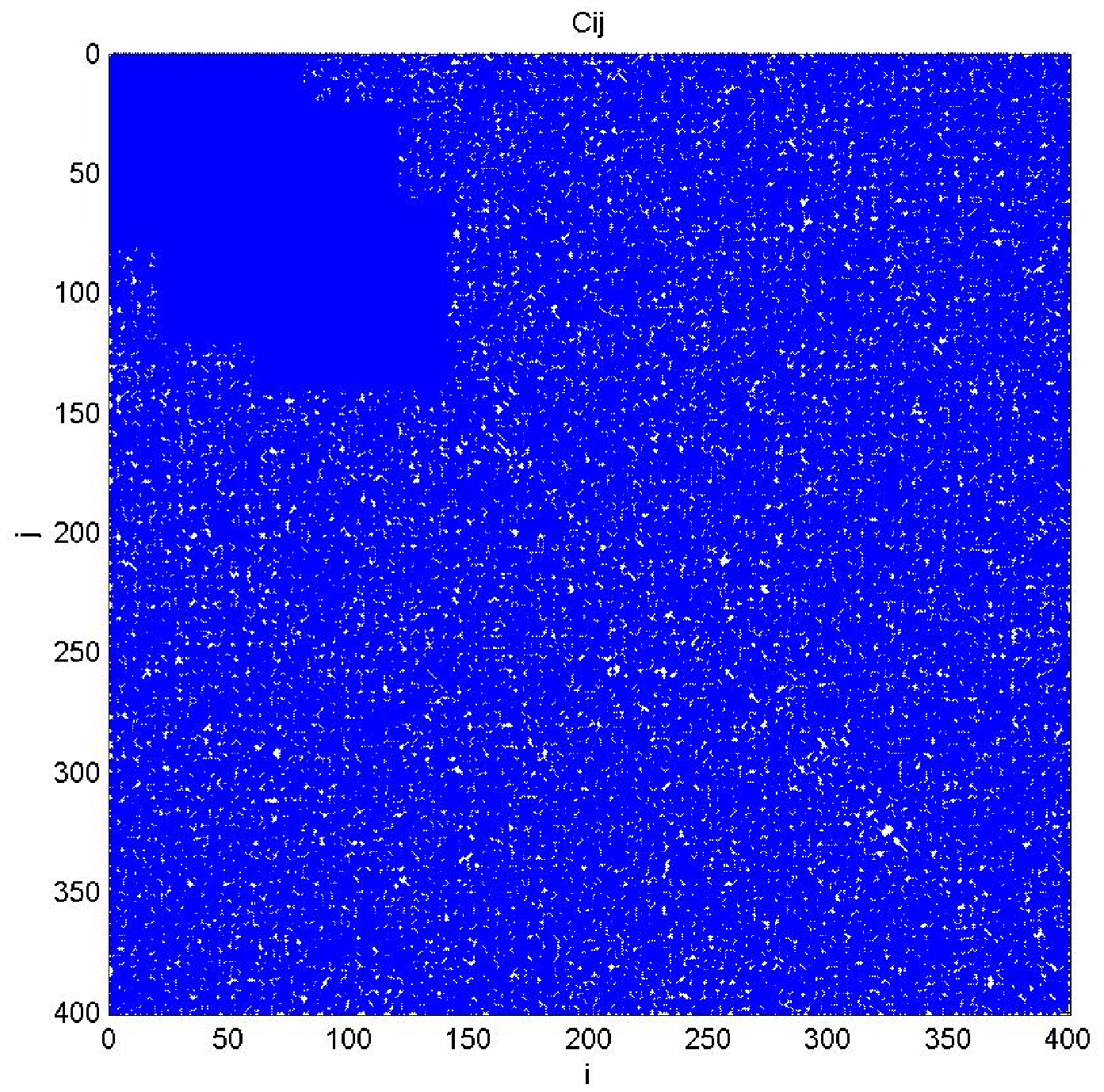}
\caption{Connectivity matrix with three overlapping cliques and 70\% random background.}
\label{fig:C3in400}
\end{figure}

\begin{figure}[h]
\includegraphics{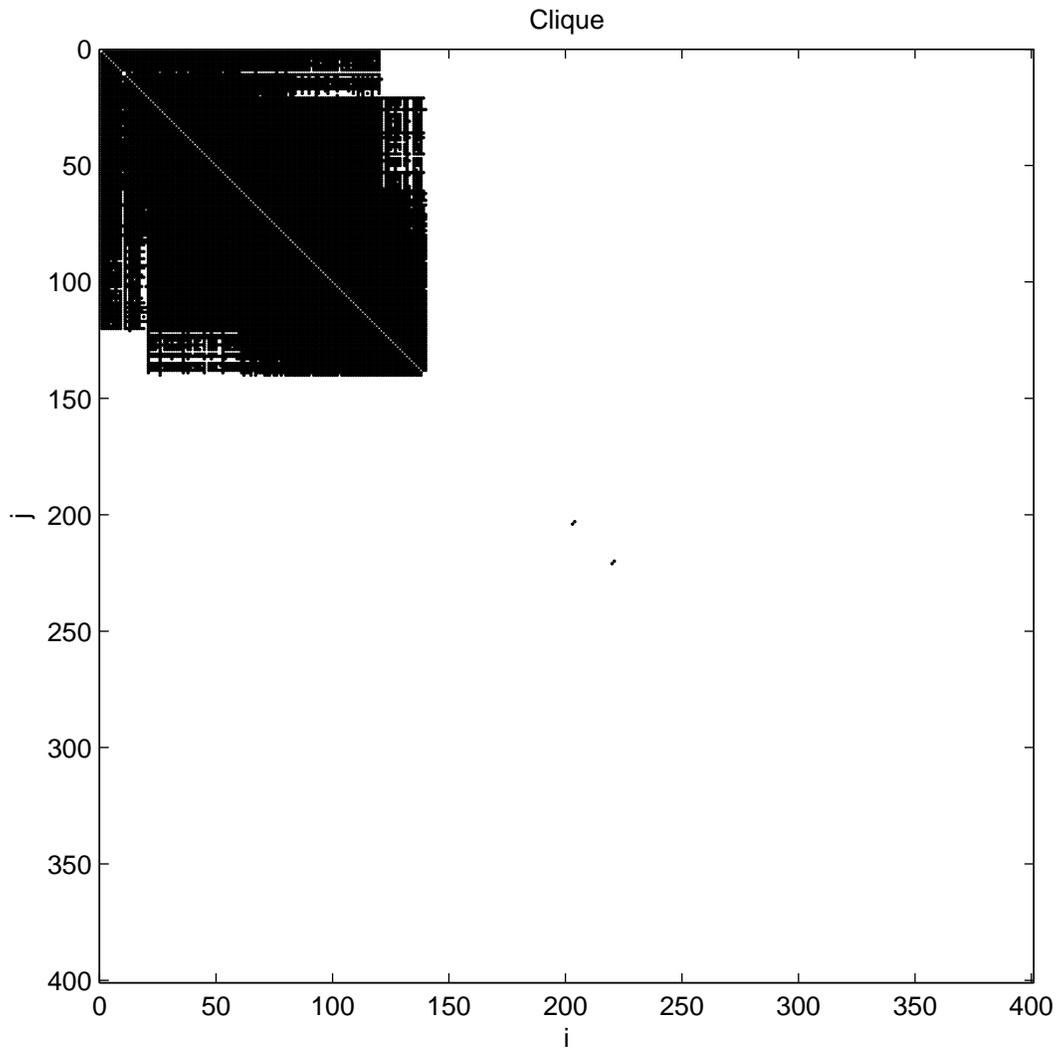}
\caption{Reconstructed cliques for 400-matrix with 70\% background connectivity.}
\label{fig:CL3in400}
\end{figure}

  There is one  ``small'' correction however to the above argument.
  It is due to the fact that in our original algorithm we have introduced one
 further constraint on the motion of the points, namely that  at all times on the unit circle
                        $|\vec{r}_i(t)| = 1$.
 It seemed necessary in order  to avoid running away to infinity of
  repelling  vertices or  collapse to the origin of  attracting ones.
   This does however introduce an extra normal reaction force that could
  in fact cancel the above sum in eq.(\ref{z}), and thus yields local minima.
  Hence in the final runs we did not impose this constraint. Instead we modified our code to facilitate handling the increasing distances between points at later stages of the evolution.  We found that our program fully reconstructed the maximal clique\cite{fn3}.  This happens regardless of the degree of the connectivity of the random background and also of the existence of large and partially overlapping slightly smaller cliques. Thus for the n=100 maximal clique in an N=400 vertex graph (i.e the students choice for dorm problem) we added two 80x80 cliques which overlapped our 100x100 clique in two 60x60 patches which ,in turn, had a 20x20 overlap and used a background with 70\% connectivity Fig.(\ref{fig:C3in400}). 
 Even under such seemingly unfavorable conditions we reconstructed our clique Fig.(\ref{fig:CL3in400}).


\begin{thebibliography}{}
\bibitem{fn1} Social pressures / encouragement via nasty / favoureble communications   are not limited by dimensionality-explaining its vast power and magically efficiency.
\bibitem{fn2}This choice common to the present and earlier works, was originally made for simplicity. In retrospect it turns to be optimal.  Indeed if the forces -  or  potentials become too strong at short distances,then two points which attract each other, may prematurely "seize" and become inseparable - a phenomenon which is a source of the deleterious multiple local minima.  The general convexity condition for the potentials $U(r)$ which avoids this difficulty -and which our linear potential trivially satisfies- has been pointed to us by Vassilios S. Vassiliadis (private
communication).
\bibitem{fn3} Once we find a substantial portion $l< n$ of the vertices belonging in the perfect clique the construction of the rest of the clique can be easily completed as follows:  We omit from our graph G all those vertices which do not connect to ALL the  vertices in the clique. This will drastically reduce the number of vertices in the remaining relevant part of G and vastly accelerate the remaining calculations.
\end{thebibliography}
\end{document}